\documentclass[11pt]{article}
\usepackage{amsfonts, amsbsy, amssymb}

\begin{document}


\newtheorem{theorem}{Theorem}
\newtheorem{proposition}{Proposition}
\newtheorem{corollary}{Corollary}
\newtheorem{lemma}{Lemma}

\def \R{{\mathbb R}}
\def \T{{\mathbb T}}
\def \Z{{\mathbb Z}}
\def \eps{\varepsilon}
\def \u {\boldsymbol u}

\title{A new integrable 3+1 dimensional generalization of the Burgers
equation  \footnotetext{AMS subject classification 35Q53, 35Q58}}
\author{M. Rudnev, A.V. Yurov, V.A. Yurov }

\maketitle

\maketitle
\begin{abstract}
A new nonlinear 3+1 dimensional evolution equation admitting the Lax pair is
presented. In the case of one spatial dimension, the equation reduces to the
Burgers equation. A method of construction of exact solutions, based on a class
of discrete symmetries of the former equation is developed. These symmetries
reduce to the Cole-Hopf transformation in one-dimensional limit. Some exact
solutions are analyzed, in the physical context of spatial dissipative
structures and shock wave dressing.
\end{abstract}

\subsection*{1. Introduction} The vast majority of known completely
integrable nonlinear evolution PDEs is 1+1 dimensional \cite{AS}, \cite{N}.
Dealing with more than a single spatial dimension, one faces fundamental
algebraic and geometric obstructions. This fact accounts for the scarcity of
integrable $d$+1 dimensional systems, for $d\geq 2$, available today.

A notable exception is the Burgers equation, which in 1+1 dimensions
is
\begin{equation}
u_t+uu_x-\nu u_{xx}=0.\label{be}
\end{equation}
It generalizes to $d\geq2$ spatial dimensions as follows
\begin{equation}
\u_t+\u\cdot\nabla \u-\nu \nabla^2
\u=0,\;\;\u=-\nabla\Phi,\;\;\nabla=(\partial_{x_1},\ldots,\partial_{x_d}).
\label{mty}
\end{equation}
from the physical point of view, the equation describes the balance between
nonlinearity and dissipation, \cite{B}, rather than dispersion, which is more
characteristic of integrable evolution equations. It comes as a model in
various problems of continuous media dynamics, condensed matter physics
cosmology, etc., see for instance \cite{FB}, \cite{HV}, \cite{P}.

Mathematically, the Burgers equation is special, as it can be fully
linearized via the Cole-Hopf (\cite{C}, \cite{H}) substitution
\begin{equation}
\u=-2\nu\nabla\log\Theta,\label{ch}
\end{equation} which reduces it to the heat equation for $\Theta$, with
diffusivity $\nu$. For the state of the art on the pure mathematical side of the Burgers equation,
see for instance \cite{WKMS} and the references contained therein.

The principal content of this note is as follows. Apart from the natural
integrable higher dimensional generalization (\ref{mty}) of the Burgers
equation (\ref{be}), the latter can be used as the foundation to construct
higher dimensional integrable PDEs, are not fully linearizable, yet their
solutions can be found via Lax pairs. In the recent work \cite{Y}, the 2+1
dimensional BLP (Boiti, Leon and Pempinelli) system was studied and shown to
reduce to an integrable two dimensional generalization of the Burgers equation
(\ref{be}). Here we present a 3+1 dimensional nonlinear equation, which
contains dissipative additives and has the following properties:
\begin{enumerate}
\item It is a scalar second order evolution PDE with quadratic nonlinearity.
\item In one dimensional limits this equation reduces to the Burgers
equation. However, unlike the $d$+1 dimensional Burgers equation,
our equation is non-isotropic, nor it is linearizable via the
Cole-Hopf substitution.
\item It allows for an explicit Lax pair.
\item It possesses a class of Darboux-transformation-like discrete symmetries, and
to take advantage of these symmetries one has to solve
the Lax pair  equations. The symmetries generate a rich spectrum of exact
solutions of the equation. In
particular they enable one to fulfill a 3+1 dimensional dressing of particular solutions of the 1+1
dimensional Burgers equation.  Conversely, in one dimensional
limits the symmetries reduce to the Cole-Hopf transformation.
\end{enumerate}

\subsection*{2. General result}
Consider the following equation:
\begin{equation}
\begin{array}{c}
\displaystyle{K[u]\equiv u_t+a_1 \left({u_x}^2-u_{xx}\right)+a_2
\left({u_z}^2-u_{zz}\right)+b_1 \left(u_x u_y-u_{xy}\right)+b_2
\left(u_x u_z-u_{xz}\right)-\rho u_x-}
\\
\displaystyle{-\mu u_y-\lambda u_z=0},
\end{array}
\label{Synok}
\end{equation}
where $u=u(t,x,y,z,t)$, and all other parameters are constants.

\medskip\noindent
In one dimensional limits equation (\ref{Synok}) reduces
to the dissipative Burgers equation. Indeed, if $u=u(x,t)$, defining
\begin{equation} \xi(x,t)=u_x(x,t),\label{oned}\end{equation} we
obtain
\begin{equation}
\xi_t-\rho\xi_x-a_1\xi_{xx}+2a_1\xi\xi_x=0. \label{Bx}
\end{equation}
The latter equation boils down to the Burgers equation after the
change $t\to t'$, such that
$$
\partial_{t'}=\partial_t-\rho\partial_x,
$$
or simply letting $\rho=0$.

In the same fashion, if $u=u(z,t)$ and $\eta(z,t)=u_z(z,t)$, one has
\begin{equation}
\eta_t-\lambda\eta_z-a_2\eta_{zz}+2a_2\eta\eta_z=0, \label{Bz}
\end{equation}
the analog of (\ref{Bx}).

Finally, the reduction $u=u(y,t)$ results in a linear equation
$$
u_t-\mu u_y=0.
$$
In view of the above, equation (\ref{Synok}) can be viewed as a special
non-isotropic
three dimensional extension of the Burgers equation. To emphasize
this, let $w=u_x$, consider $\mu=\rho=\lambda=0$ and rewrite
(\ref{Synok}) as follows:
\begin{equation}
\begin{array}{c}
w_t+2a_1ww_x+b_1ww_y+b_2ww_z-a_1w_{xx}-a_2w_{zz}-b_1w_{xy}-b_2w_{xz}-\mu
w_y+\\
+b_1u_yw_x+b_2u_zw_x+2a_2u_zw_z=0.
\end{array}
\label{Synok1}
\end{equation}

Our main result is the following theorem
\begin{theorem} \label{th}Let $u(x,y,z,t)$  be a particular solution of equation (\ref{Synok}) and
 $\psi=\psi(x,y,z,t)$ be a solution of the following linear equation:
\begin{equation}
\begin{array}{c}
\displaystyle{ \psi_t=a_1 \psi_{xx}+a_2 \psi_{zz}+b_1
\psi_{xy}+b_2 \psi_{xz}+\left(\rho-2 a_1 u_x-b_2 u_z-b_1
u_y\right) \psi_x+\left(\mu-b_1 u_x\right) \psi_y+}
\\
\displaystyle{+\left(\lambda-2 a_2 u_z-b_2 u_x\right) \psi_z}\equiv
{\boldsymbol A}[u]\, \psi.
\end{array}
\label{A}
\end{equation}
Then any ${\tilde u}_{klm} ={\tilde u}_{klm}(x,y,z,t),$ defined by
the formula
\begin{equation}
{\tilde u}_{klm}=u-\log\left(
\left(\partial_x-u_x\right)^{k}\left(\partial_y-u_y\right)^{l}
\left(\partial_z-u_z\right)^{m}\psi\right),
\;\;(k,l,m)\in\Z^3_+\label{result}
\end{equation}
is also a solution of equation (\ref{Synok}).
\end{theorem}
\addtocounter{proposition}{1}
\addtocounter{corollary}{2}
Theorem \ref{th} rests on the following fact.
\begin{proposition}
Equation (\ref{Synok}) admits the following Lax pair:
$\psi_t={\boldsymbol A}(u)\psi$, cf. (\ref{A}), and
\begin{equation}
\begin{array}{c}
\displaystyle{
\psi_{xyz}=u_z\psi_{xy}+u_y\psi_{xz}+u_x\psi_{yz}+\left(u_{yz}-u_yu_z\right)\psi_x+\left(u_{xz}-u_xu_z\right)\psi_y+
\left(u_{xy}-u_xu_y\right)\psi_z+}
\\
\displaystyle{\left(u_{xyz}-u_{yz}u_x-u_{yx}u_z-u_{xz}u_y+u_xu_yu_z\right)\psi},
\end{array}
\label{L}
\end{equation}
\end{proposition}

Verification of this proposition a direct calculation. Further in
the note we shall refer to equations (\ref{L}) and (\ref{A}) as the
Lax, or LA-pair for equation (\ref{Synok}), and $u$ as a potential.
Observe that the spectral equation (\ref{L}) of the Lax pair can be
rewritten in a more compact form:
\begin{equation}
\left(\partial_x-u_x\right)\left(\partial_y-u_y\right)
\left(\partial_z-u_z\right)\psi\equiv{\boldsymbol L}_1[u]{\boldsymbol L}_2[u]{\boldsymbol
L}_3[u]\psi\equiv{\boldsymbol L}[u]\psi=0. \label{factor}
\end{equation}
Also observe that if we redefine the operator ${\boldsymbol A}[u]\to {\boldsymbol
A}'[u]= {\boldsymbol A}[u]+K[u]$, then the compatibility condition of the
Lax pair equations (\ref{L}) and (\ref{A}) will be reduced to a
an identity. Namely the operators $\boldsymbol L[u]$ and ${\boldsymbol
B}'[u]=\partial_t-{\boldsymbol A}'[u]$ will commute:
$[\boldsymbol{L}[u],\boldsymbol{B}'[u]]=0$.

Theorem \ref{th} implies the following corollary, which follows after
successive iteration of (\ref{result}).
\begin{corollary} If $\{\psi_i\}$, $i=1,..,N$ is a set of particular solutions
of the Lax pair  (\ref{L}), (\ref{A}), given the potential $u$,
satisfying equation (\ref{Synok}), new solutions of (\ref{Synok})
are generated by the following rule:
\begin{equation}
{\tilde u} =u-\log\left(\prod_{i=1}^N   {\boldsymbol L}^{k_i}_1[u]{\boldsymbol
L}^{l_i}_2[u]{\boldsymbol L}^{m_i}_3[u]\psi_i
\right),\;\;\;(k_i,l_i,m_i)\in\Z^3_+,\;\forall i=1,\ldots,N.
\label{cr}\end{equation}
\end{corollary}

As we have indicated earlier, the Burgers equation
(\ref{Bx}) results from a 1+1 dimensional reduction of
equation (\ref{Synok}). Conversely, formulae (\ref{result}),
(\ref{cr}) yield a bona fide generalization of the Cole-Hopf
substitution (\ref{ch}). Indeed,  setting $u\equiv 0$, $k=l=m=0$ in (\ref{result}), after differentiation in $x$, we obtain
the one dimensional Cole-Hopf transformation, cf. (\ref{ch}). Clearly, the
same can be said about the $z$ variable reduction as well.

\subsection*{Proof of Theorem \ref{th}}

The theorem will follow from the following lemma.
\addtocounter{lemma}{3}
\begin{lemma} Let $\psi$ be a solution of the Lax pair equaitons (\ref{L}), (\ref{A}) with
the potential $u$, which  is a solution of equation (\ref{Synok}). Then the
function
\begin{equation}
{\tilde \psi}_{klm}={\boldsymbol L}^{k}_1[u]{\boldsymbol L}^{l}_2[u]{\boldsymbol
L}^{m}_3[u]\psi, \;\;\;(k,l,m)\in\Z^3_+\label{Lemma}
\end{equation}
also satisfies (\ref{L}), (\ref{A}), with the same
potential $u$.\label{lm}
\end{lemma}
To prove the lemma, observe the validity of the following
commutator relations, for $i,j=1,2,3$:
$$
[{\boldsymbol L}_i[u],{\boldsymbol L}[u]]=[{\boldsymbol L}_i[u],{\boldsymbol B}[u]]=[{\boldsymbol
L}_i[u],{\boldsymbol L}_j[u]]=0.
$$
Lemma \ref{lm} then follows by induction. $\Box$

Now, to prove Theorem \ref{th}, let us introduced
three intertwining operators
\begin{equation}
{\boldsymbol D}_i=f_i\partial_i-g_i,\qquad i=1,2,3 \label{D},
\end{equation}
with the quantities $f_i, g_i$ to be found (naturally, $\partial_{1,2,3}=\partial_{x,y,z}$,
respectively), such that
the operators
${\boldsymbol D}_i$ have the following property: for some $u_i=u_i(x,y,z,t),$
\begin{equation}
{\boldsymbol L}(u_i){\boldsymbol D}_i={\boldsymbol D}_i {\boldsymbol L}[u],\qquad {\boldsymbol B}(u_i){\boldsymbol
D}_i={\boldsymbol D}_i {\boldsymbol B}[u]. \label{spl}
\end{equation}
The commutation relations (\ref{spl}) determine the maps $u\to u_i$,
which come from substitution of (\ref{D}) into
(\ref{spl}). The explicit form of the operators ${\boldsymbol D}_i$ is found
as follows.

Substituting  (\ref{D}) into (\ref{spl}) and equating the components
at the same partial derivatives results in a system of nonlinear
equations (which is not quoted because of its bulk) whence it follows:
\begin{equation}
{\boldsymbol D}_i={\rm e}^{-v}\left({\boldsymbol L}_i[u]-c_i\right),\qquad
u_i={\tilde u}=u-v, \label{new}
\end{equation}
where $c_i$ - are constants, which will be further assigned zero values.
The quantity $v=v(x,y,z,t)$ is a solution of the following nonlinear equation:
\begin{equation}
\begin{array}{c}
\displaystyle{v_t=a_1 \left(v_{xx}+{v_x}^2\right)+a_2
\left(v_{zz}+{v_z}^2\right)+b_1 \left(v_{xy}+v_x v_y\right)+b_2
\left(v_{xz}+v_x v_z\right)+}
\\
\displaystyle{+\left(\rho-2 a_1 u_x-b_2 u_z-b_1 u_y\right)
v_x+\left(\mu-b_1 u_x\right)v_y+\left(\lambda-2 a_2 u_z-b_2
u_x\right)v_z}
\end{array}
\label{vt}
\end{equation}
Therefore (\ref{new}) or explicitly (\ref{vt}) indicate that for
$u\equiv 0,$ the function "$-v$" satisfies equation (\ref{Synok}).

Then automatically the quantity $\psi={\rm e}^{-u}$ will satisfy the
Lax pair equations (\ref{L}) and (\ref{A}). In fact, the L-equation,
cf. (\ref{L}), is satisfied as the identity. The A-equation
(\ref{A}) however is satisfied only if $u$ is a solution of
(\ref{Synok}). On the other hand, by Lemma \ref{lm}, the functions of the
form ${\tilde \psi}_{klm}$ defined via relation (\ref{Lemma}) are
also solutions of (\ref{L}) and (\ref{A}), with the same potential
$u$. Rewriting them as ${\tilde \psi}_{klm}=\exp(v_{klm})$ and
substituting into (\ref{A}) one verifies that the quantities
$v_{klm}$ are indeed solutions of equation (\ref{vt}). Theorem \ref{th} and
formula (\ref{result}) now follow from the second relation from
(\ref{new}). $\Box$

\medskip
\noindent {\em Remark.} Formula (\ref{result}) has a countenance
similar to the Darboux transformation, which is a standard tool for
construction of exact solutions of nonlinear PDEs (usually 1+1, more
rarely 2+1 dimensional) which admit Lax pairs, see e.g. \cite{Salle}
for the general theory, applications and references. However
(\ref{result}) does not represent a bona fide Darboux transform for
two following reasons.
\begin{enumerate}
\item Darboux transforms, representing discrete symmetries of a Lax
pair, possess a non-trivial kernel on the solution space of the
pair. In other words, there always exists some Lax pair solution
which zeroes the transform. This is the property which enabled one
Crum, \cite{Crum}, to write down the determinant formulae for
successive Darboux transforms. Transformation (\ref{result}) however
does not have this property.

It is known that in addition to the Darboux transform, fairly rich
spectral problems, such as the Zakharov-Shabat problem for the
Nonlinear Schr\"odinger equation or its two-dimensional
generalization for the Davey-Stewartson equations, admit another
discrete symmetry, namely the  Schlesinger transform, \cite{Schl}.
The difference between (\ref{result}) and the latter transformations
lies in the fact that for the Schlesinger transform, the potential
transformation rules can be locally defined without using the Lax
pair solution $\psi$, while (\ref{result}) certainly does so. This
feature is shared by (\ref{result}) and the standard Darboux
transformation.

\item To construct exact solutions of nonlinear PDEs via the Darboux
transform, one has to take advantage of the solution of the full Lax pair
as a system of equations. In order to get (\ref{result}) however, we
have used the solution of the A-equation (\ref{A}) only. The
L-equation (\ref{L}) has been used only as a tool to prove Theorem \ref{th}. To this effect, transformation (\ref{result}) combines essential
features of the Darboux and Cole-Hopf transformations.
\end{enumerate}

\subsection*{4. Some exact solutions}

Let us use the above formalism in order to construct some exact solutions
of equation (\ref{Synok1}) (which is the equation for $u_x$, where $u$ is a solution of equation
(\ref{Synok}) with $\mu=\lambda=\rho=0$). We consider equation
(\ref{Synok1}), because it appears to be a closer relation of the Burgers equation and is likely to
be interesting from the physics point of view.

{\bf Example 4.1.} As the first example let us consider dressing on the vacuum
background $u\equiv 0$. In this case the function
\begin{equation}
\psi(x,y,z,t)=a^2x^2+b^2y^2+c^2z^2+2\left(a_1a^2+a_2c^2\right)t+s^2,
\label{1}
\end{equation}
where $a,b,c,s$ are some real constants, is clearly a solution of the Lax pair
equations (\ref{L}) and (\ref{A}). Substituting (\ref{1}) in (\ref{result}) we
derive the ${\tilde u}_{klm}$. After differentiating it with respect to $x$ and
choosing $k=l=m=0$, we obtain a solution $w$ of equation (\ref{Synok1}) as
follows:
\begin{equation}
w(x,y,z,t)=-\frac{2a^2x}{a^2x^2+b^2y^2+c^2z^2+2\left(a_1a^2+a_2c^2\right)t+s^2}.
\label{1.1}
\end{equation}
Physically, solution (\ref{1.1}) describes a rationally localized impulse,
vanishing as $t\to +\infty$. To ensure that (\ref{1.1}) is non-singular for
$t\ge 0$, one should impose the inequality $a_2\ge -a^2a_1/c^2$ on the
coefficients. Moreover, if
$a_2= -a^2a_1/c^2$, the solution in question is rationally localized and stationary.

{\em Remark.} The fact that there exists a localized stationary
solution in an equation containing dissipative terms may appear
somewhat paradoxical from the point of view of physics. However solution
(\ref{1.1}) is stationary only if $a_1a_2<0$.  One can see that
these constants appear in the dissipative terms of equation
(\ref{Synok1}), $a_1$ characterizing the dissipation along the $x$
and $a_2$ along the $z$ axes. The fact that $a_1$ and $a_2$ should
have different signs implies that dissipation in the direction of
one axis is compensated by instability in the direction of
the other. The balance of these two effects results in the
stationary solution, which can be regarded as a three-dimensional
{\em dissipative structure}. A similar situation occurs with
two-dimensional stationary solutions of the BLP equation, cf.
\cite{Y}.

\medskip
{\bf Example 4.2.} Another solution of (\ref{L}), (\ref{A}) when $u\equiv 0$ is
\begin{equation}
\psi(x,y,z,t)=c_1{\rm  e}^{\alpha(\alpha a_1+\beta b_1)t}\cosh(\alpha
x+\beta y)+c_2 {\rm
e}^{(a_1a^2+a_2b^2+abb_2)t}\cosh(ax+bz)+c_3{\rm
e}^{a_2c^2t}\cosh(cz), \label{2}
\end{equation}
where $\alpha,\beta, a,b,c_1,c_2,c_3$ are some real
constants. Choosing them such that
\begin{equation}
\beta=-\frac{\alpha a_1}{b_1},\qquad
b=\frac{-b_2\pm\sqrt{b_2^2-4a_1a_2}}{2a_2}ю\label{uh}
\end{equation}
and using (\ref{2}) and (\ref{uh}) in the same way as in Example 4.1 above (\ref{1.1}), we
obtain another solution:
\begin{equation}
\displaystyle{ w(x,y,z,t)=-\frac{\alpha c_1\sinh(\alpha x+\beta y)+a
c_2\sinh(ax+bz)}{c_1\cosh(\alpha x+\beta
y)+c_2\cosh(ax+bz)+c_3{\rm e}^{a_2c^2t}\cosh(cz)}}. \label{2.1}
\end{equation}

\medskip
{\bf Example 4.3.} Developing the analog with \cite{Y} further, one
can develop a procedure of construction of exact solutions of the
three-dimensional equation (\ref{Synok1}) which are based initially
on the solutions of the 1+1 dimensional Burgers equation. Consider
equation (\ref{Bx}) for the unknown $\xi$. Suppose $\lambda=\mu=\rho=0$, let $a_1=\nu$ in
(\ref{Bx}).

Clearly the quantity $U(x,t)=2\nu\xi(x,t)$ solves the one
dimensional Burgers equation
\begin{equation}
U_t+UU_x-\nu U_{xx}=0. \label{Burg}
\end{equation}
As a starting point let us take a shock wave solution of (\ref{Burg}), e.g.
\begin{equation}
\xi=U_x=\frac{v-\nu a}{2\nu}+\frac{a}{1+{\rm e}^{a(x-vt)}},
\label{I}
\end{equation}
where  $a$ and $v$ are constants. Seek a solution of (\ref{A}) as a
superposition
\begin{equation}
\psi=\sum_{k=1}^N A_k(\eta){\rm e}^{\beta_k y+\gamma_k z},
\label{II}
\end{equation}
where $\eta=x-vt$,  while the $2N$ quantities $\beta_k$ and $\gamma_k$
can in general be functions of $\eta$. For simplicity however let us render
them constants to be determined.

Substitution of (\ref{II}) into (\ref{A}) yields $N$ linear equations
for $A_k(\eta)$:
\begin{equation} \nu{\ddot
A}+\left(v+\sigma-2\nu\xi\right){\dot
A}+\left(a_2\gamma^2-\sigma\xi\right)A=0. \label{III}
\end{equation}
In equation (\ref{III}) the indices for the quantities $A_k$,
$\beta_k$, $\gamma_k$ and $\sigma_k\equiv b_1\beta_k+b_2\gamma_k$
have been omitted, while the quantity $\xi$ came from (\ref{I}), dot
standing for differentiation with respect to $\eta$.

Equation (\ref{III}) can be simplified further. To do so, let us
introduce a new independent variable:
$$
q=\xi(x,t)-\xi_0,\qquad {\rm e}^{a\eta}=\frac{a}{q}-1,\qquad
\xi_0=\frac{v-\nu a}{2\nu}.
$$
In terms of $q$ equation (\ref{III}) becomes
\begin{equation}
\nu\left(q^2-aq\right)^2A''(q)+\sigma\left(q^2-aq\right)A'(q)+\left(\delta-\sigma
q\right)A(q)=0, \label{IV}
\end{equation}
where $\delta=a_2\gamma^2-\sigma\xi_0$, while prime stands for
differentiation in $q$. The latter equation can be
simplified further via a substitution
\begin{equation}
A(q)=W(q)\left(\frac{q}{q-a}\right)^{\sigma/(2\nu a)}, \label{V}
\end{equation}
reducing (\ref{IV}) to the following equation for $W(q)$:
\begin{equation}
\frac{W''(q)}{W(q)}=\frac{2a\sigma\nu-4\delta\nu+\sigma^2}{4\nu^2(q-a)^2q^2}.
\label{VI}
\end{equation}
In a particular case of the dependence
$$
\beta_k=\frac{-b_2\gamma_k-\nu\pm\sqrt{\nu^2+4a_2\nu\gamma_k^2}}{b_1},
$$
one can solve (\ref{A})  explicitly as
\begin{equation}
\psi=\sum_{k=1}^N\left(W_kq+V_k\right)
\left(\frac{q}{q-a}\right)^{\sigma_k/(2\nu a)}{\rm
e}^{\beta_ky+\gamma_kz}, \label{last}
\end{equation}
where $W_k$ and $V_k$ are arbitrary constants. Substitution of
(\ref{last}) into (\ref{result}) yields exact solutions of
equations (\ref{Synok}), (\ref{Synok1}). The described procedure
can be regarded as the shock wave dressing.

\vspace{.2in} \noindent{\large {\bf Acknowledgement:}} Research has
been partially supported by EPSRC Grant GR/S13682/01.

\medskip
\noindent {\bf Authors:}\\
{\tt Mischa Rudnev:} Department of Mathematics, University of
Bristol, Bristol BS6 6AL UK; e-mail {\em m.rudnev@bris.ac.uk}

\medskip \noindent
{\tt Artem V. Yurov:} Department of Theoretical Physics, Kaliningrad
State University, Aleksandra Nevskogo 14, Kaliningrad 236041,
Russia; e-mail {\em artyom\_yurov@mail.ru}

\medskip \noindent
{\tt Valerian A. Yurov:} Department of Theoretical Physics,
Kaliningrad State University, Aleksandra Nevskogo 14, Kaliningrad
236041, Russia; e-mail {\em yurov@freemail.ru}

\end{document}